\documentclass{article}

\usepackage{arxiv}
\newcommand{\BibDir}{./bib}
\usepackage[utf8]{inputenc} 
\usepackage[T1]{fontenc}    
\usepackage{hyperref}       
\usepackage{url}            
\usepackage{booktabs}       
\usepackage{amsfonts}       
\usepackage{nicefrac}       
\usepackage{microtype}      
\usepackage{amsmath}
\usepackage{lipsum}
\usepackage{graphicx}
\usepackage{caption}
\usepackage{subcaption}
\usepackage{multirow}
\usepackage{amsmath}
\newtheorem{Example}{Ex.}[section]

\title{Extracting Clinical Concepts from User Queries}

\author{
  Yue Zhao\thanks{Corresponding author.} \\
  Goergen Institute for Data Science\\
  University of Rochester\\
  Rochester, NY 14627 \\
  \texttt{yuezhao@rochester.edu} \\
   \And
  John Handley\\
  Goergen Institute for Data Science\\
  University of Rochester\\
  Rochester, NY 14627 \\
  \texttt{john.handley@rochester.edu} \\
}

\begin{document}
\maketitle

\begin{abstract}
Clinical concept extraction often begins with clinical Named Entity Recognition (NER). Although a number of annotated clinical notes are available, NER models trained exclusively on the sentences from the annotated clinical notes tend to struggle with tagging clinical entities in user queries because of the structural differences between clinical note sentences and user queries. In many cases, for example, user queries are compounded of multiple clinical entities, without comma or conjunction words separating them. By using as dataset a mixture of annotated clinical note sentences and synthesized user queries which require no manual annotation, we adapt a clinical NER model based on the BiLSTM-CRF architecture for tagging clinical entities in user queries. Our contribution are the following: 1) We found that when trained on a mixture of synthesized user queries and clinical note sentences, the NER model performs better on both types of input data. 2) We provide an end-to-end and easy-to-implement framework for clinical concept extraction from user queries. 
\end{abstract}

\section{Introduction}\label{Intro}
Medical search engines are an essential component for many online medical applications, such as online diagnosis systems and medical document databases. A typical online diagnosis system, for instance, relies on a medical search engine. The search engine takes as input a user query that describes some symptoms and then outputs clinical concept entries that provide relevant information to assist in diagnosing the problem. One challenge medical search engines face is the segmentation of individual clinical entities. When a user query consists of multiple clinical entities, a search engine would often fail to recognize them as separate entities. For example, the user query “fever joint pain weight loss headache” contains four separate clinical entities: “fever”, “joint pain”, “weight loss”, and “headache”. But when the search engine does not recognize them as separate entities and proceeds to retrieve results for each word in the query, it may find "pain" in body locations other than "joint pain", or it may miss "headache" altogether, for example. Some search engines allow the users to enter a single clinical concept by selecting from an auto-completion pick list. But this could also result in retrieving inaccurate or partial results and lead to poor user experience. 

We want to improve the medical search engine so that it can accurately retrieve all the relevant clinical concepts mentioned in a user query, where relevant clinical concepts are defined with respect to the terminologies the search engine uses. The problem of extracting clinical concept mentions from a user query can be seen as a variant of the Concept Extraction (CE) task in the frequently-cited NLP challenges in healthcare, such as 2010 i2b2/VA  \cite{i2b2_2010} and 2013 ShARe/CLEF Task 1 \cite{Suominen}. Both CE tasks in 2010 i2b2/VA and 2013 ShARe/CLEF Task 1 ask the participants to design an algorithm to tag a set of predefined entities of interest in clinical notes. These entity tagging tasks are also known as clinical Named Entity Recognition (NER). For example, the CE task in 2010 i2b2/VA defines three types of entities: “problem”, “treatment”, and “test”. The CE task in 2013 ShARe/CLEF defines various types of disorder such as “injury or poisoning”, "disease or syndrome”, etc. In addition to tagging, the CE task in 2013 ShARe/CLEF has an encoding component which requires selecting one and only one Concept Unique Identifier (CUI) from Systematized Nomenclature Of Medicine Clinical Terms (SNOMED-CT) for each disorder entity tagged. Our problem, similar to the CE task in 2013 ShARe/CLEF, also contains two sub-problems: tagging mentions of entities of interest (entity tagging), and selecting appropriate terms from a glossary to match the mentions (term matching). However, several major differences exist. First, compared to clinical notes, the user queries are much shorter, less technical, and often less coherent. Second, instead of encoding, we are dealing with term matching where we rank a few best terms that match an entity, instead of selecting only one. This is because the users who type the queries may not have a clear idea about what they are looking for, or could be laymen who know little terminology, it may be more helpful to provide a set of likely results and let the users choose. Third, the types of entities are different. Each medical search engine may have its own types of entities to tag. There is also one minor difference in the tagging scheme between our problem and the CE task in 2013 ShARe/CLEF - We limit our scope to dealing with entities of consecutive words and not disjoint entities \footnote{See section 1.8 of annotation guidelines, \url{https://drive.google.com/file/d/0B7oJZ-fwZvH5VmhyY3lHRFJhWkk/edit}}. We use only Beginning, Inside, Outside  (BIO) tags. Given the differences listed above, we need to customize a framework consisting of an entity tagging and term matching component for our CE problem.

\section{Related Work}
\label{sec:Related Work}
An effective model that has been commonly used for NER problem is a Bi-directional LSTM with a Conditional Random Field (CRF) on the top layer (BiLSTM-CRF), which is described in the next section. Combining LSTM’s power of representing relations between words and CRF’s capability of accounting for tag sequence constraints, Huang et al. \cite{Huang} proposed the BiLSTM-CRF model and used handcrafted word features as the input to the model. Lample et al. \cite{Lample} used a combination of character-level and word-level word embeddings as the input to BiLSTM-CRF. Since then, similar models with variation in types of word embeddings have been used extensively for clinical CE tasks and produced state-of-the-art results \cite{Zhu,florez,Habibi,Chalapathy}. Word embeddings have become the cornerstone of the neural models in NLP since the famous Word2vec \cite{Mikolov} model demonstrated its power in word analogy tasks. One well-known example is that after training Word2vec on a large amount of news data, we can get word relations such as $vector('king') - vector('queen') + vector('woman')  \approx vector('man')$. More sophisticated word embedding technique emerged since Word2vec. It has been shown empirically that better quality in word embeddings leads to better performance in many downstream NLP including entity tagging \cite{si,McCann}. Recently, contextualized word embeddings generated by deep learning models, such as ELMo \cite{Peters}, BERT \cite{Devlin}, and Flair \cite{akbik-flair-embeddings}, have been shown to be more effective in various NLP tasks. In our project, we make use of a fine-tuned ELMo model \footnote{\url{https://allennlp.org/elmo}} and a fine-tuned Flair model \footnote{\url{https://github.com/zalandoresearch/flair/pull/519}} in the medical domain. We experiment with the word embeddings from the two fine-tuned models as the input to the BiLSTM-CRF model separately and compare the results. 

Tang et al. \cite{Tang} provided  straightforward algorithm for term matching. The algorithm starts with finding candidate terms that contain \textit{ALL} the entity words, with term frequency - inverse document frequency (tf-idf) weighting. Then the candidates are ranked based on the pairwise cosine distance between the word embeddings of the candidates and the entity.

\section{Framework}
\label{sec:Framework}
We adopt the tagging - encoding pipeline framework from the CE task in 2013 ShARe/CLEF. We first tag the clinical entities in the user query and then select relevant terms from a glossary in dermatology \footnote{We worked with \href{https://www.visualdx.com/}{visualDx} on this project and used their glossary. See Acknowledgment.} to match the entities. 

\subsection{Entity Tagging}

\begin{figure*}[t]
\centering
\includegraphics[width=0.7\textwidth]{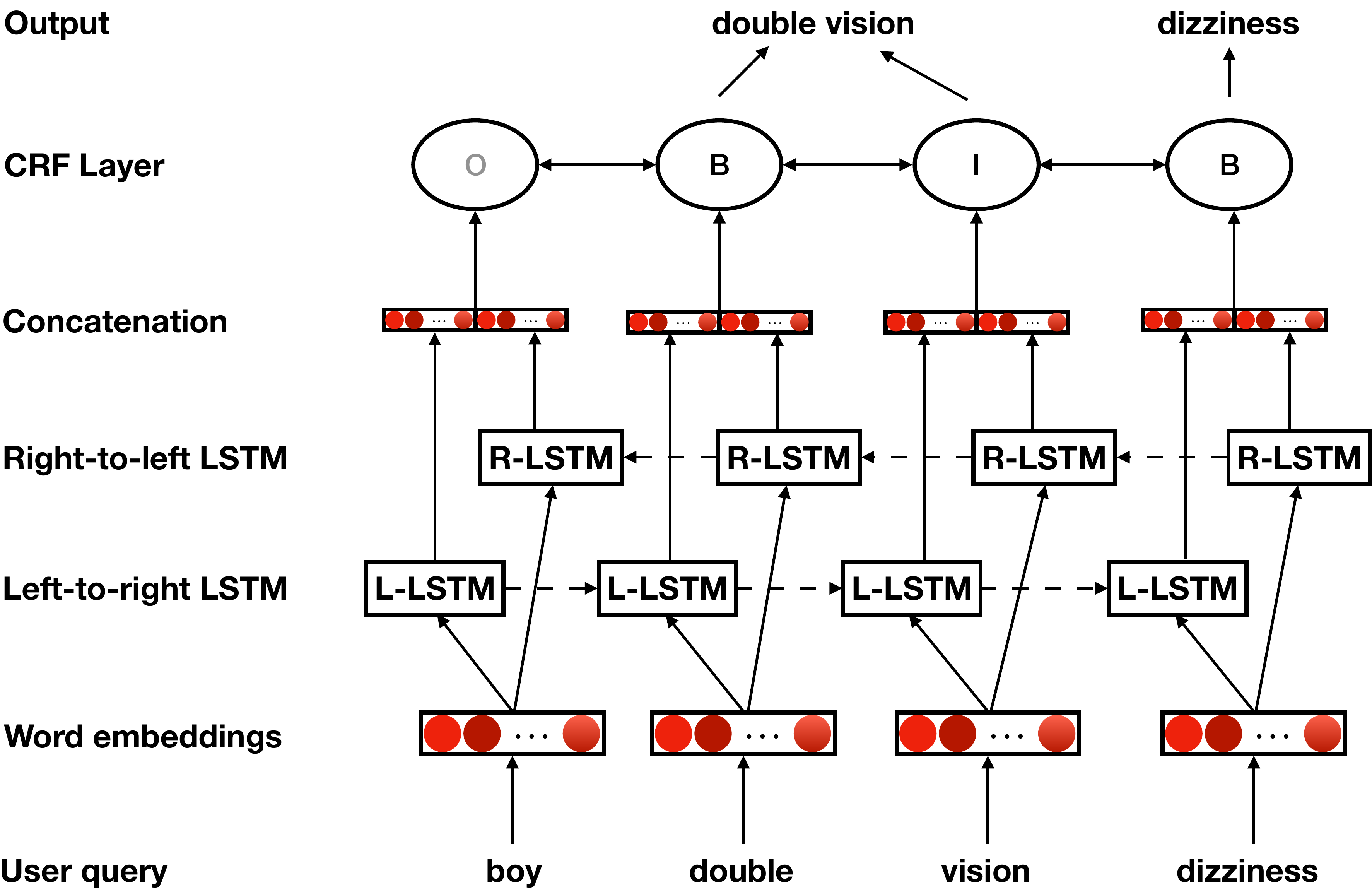}
\caption{The BiLSTM-CRF model for tagging entities in user queries. A user query, "boy double vision dizziness", enters at the bottom level. The word embeddings, shown as arrays of red dots which each represents a real number, are generated by the fine-tuned ELMo model and then fed into a left-to-right LSTM and a right-to-left LSTM. The scores output by the two LSTMs are concatenated to form the word features for the CRF layer. The CRF layer finds the most likely BIO tags for each word based on the word features and tag sequence constraints between tags. The output is the clinical entities in the query, "double vision" and "dizziness". Figure after Jurafsky \& Martin \cite{Jurafsky}. }
\label{fig:BiLSM-CRF}
\end{figure*}

We use the same BiLSTM-CRF model proposed by Huang et al. \cite{Huang}. An illustration of the architecture is shown in Figure \ref{fig:BiLSM-CRF} . Given a sequence (or sentence) of n tokens, $r = (w_1, w_2,..., w_n)$, we use a fine-tuned ELMo model to generate contextual word embeddings for all the tokens in the sentence, where a token refers to a word or punctuation. We denote the ELMo embedding, $x$, for a token $w$ in the sentence $r$ by $x = ELMo(w|r)$. The notation and the procedure described here can be adopted for Flair embeddings or other embeddings. Now, given a sequence of tokens in ELMo embeddings, $X =(x_1, x_2, ..., x_n)$, the BiLSTM layer generates a matrix of scores, $P(\theta)$ of size $n \times k$, where $k$ is the number of tag types, and $\theta$ is the parameters of the BiLSTM. To simplify notation, we will omit the $\theta$ and write $P$. Then, $P_{i,j}$ denotes the score of the token, $x_i$, being assigned to the $j$th tag. Since certain constraints may exist in the transition between tags, an "O" tag should not be followed by an "I" tag, for example, a transition matrix, $A$, of dimension $(k+2)\times(k+2)$, is initialized to model the constraints. The learnable parameters, $A_{i,j}$, represent the probability of the $j$th tag follows the $i$th tag in a sequence. For example, if we index the tags by: 1:“B”, 2:“I”, and 3:“O”, then $A_{1,3}$ would be the probability that an “O” tag follows a “B” tag. A beginning transition and an end transition are inserted in $A$ and hence $A$ is of dimension $(k+2)\times(k+2)$. 

Given a sequence of tags, $Y=(y_1,y_2,...,y_n)$, where each $y_i$, $1\leq i \leq n$, corresponds to an index of the tags, the score of the sequence is then given by
\begin{equation}\label{eq1}
score(X,Y)=\sum_{i=0}^{n} A_{y_i,y_{i+1}} + \sum_{i=1}^{n} P_{i,y_i}.    \end{equation}
The probability of the sequence of tags is then calculated by a softmax,
\begin{equation}\label{eq2}
P(Y|X)=\frac{exp(score(X,Y))}{\sum_{Y\in \{Y_x\}}exp(score(X,Y))},    
\end{equation}
where $\{Y_x\}$ denotes the set of all possible tag sequences. During training, the objective function is to maximize $\log(P(Y|X))$ by adjusting $A$ and $P$.

\subsection{Term Matching}

\begin{figure*}[t]
\centering
\includegraphics[width=0.7\textwidth]{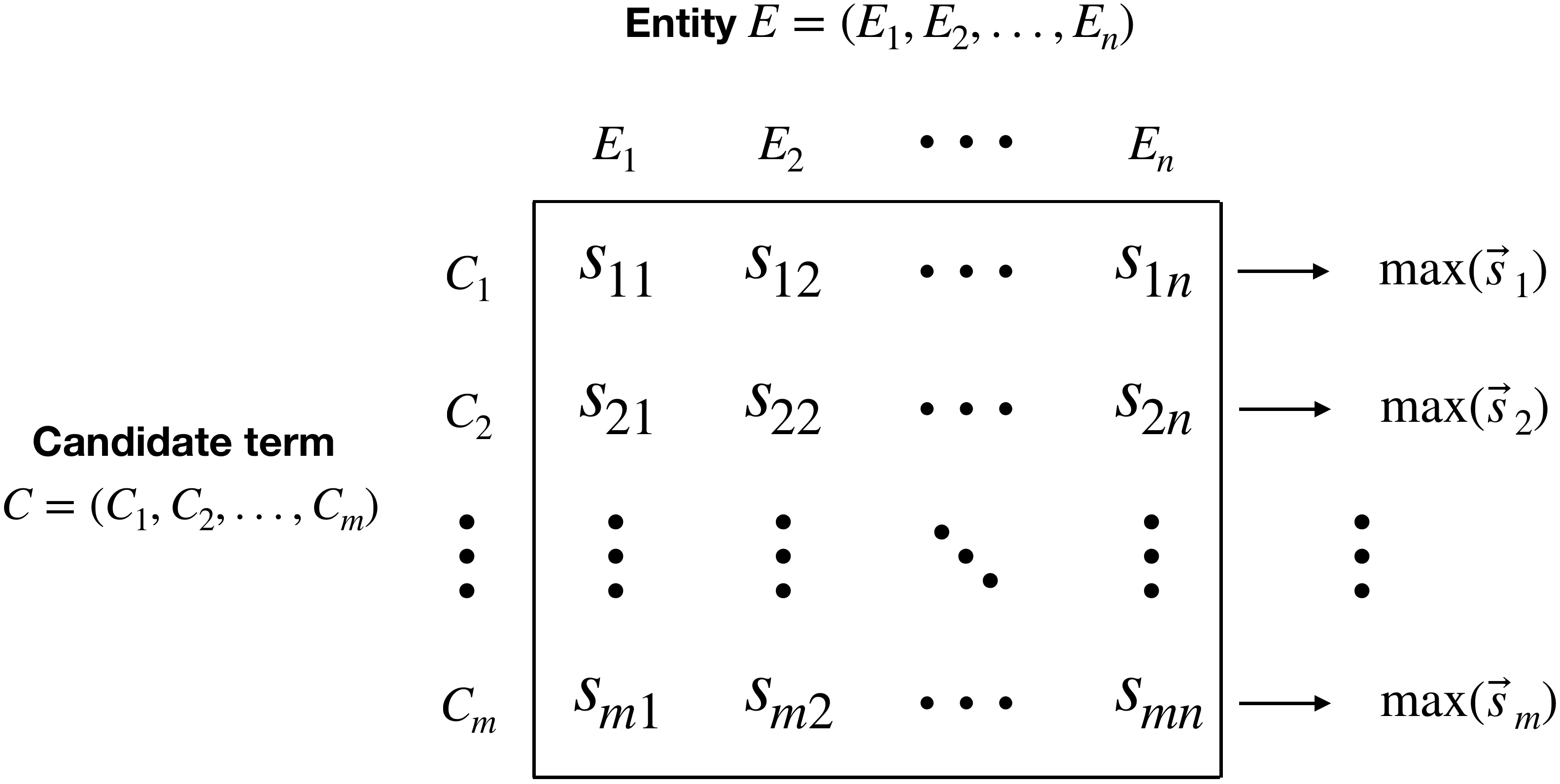}
\caption{First step in the computation for the similarity score between $E$ and $C$. A matrix of similarity scores between each word in $C$ and each word in $E$ is created. Then, for each word in $C$, get the maximum similarity score, $max(\vec{s_i})$, $1 \leq i \leq m$, where $m$ is the number of words in $C$.}
\label{fig:encoding}
\end{figure*}

The term matching algorithm of Tang et al. \cite{Tang} is adopted with some major modifications. First, to identify candidate terms, we use a much looser string search algorithm where we stem the entity words with snowball stemmer \footnote{\url{https://kite.com/python/docs/nltk.SnowballStemmer}} and then find candidate terms which contain \textit{ANY} non-stopword words in the entity. The stemming is mainly used to represent a word in its stem. For example, “legs” becomes “leg”, and “jammed” becomes “jam”. Thus, stemming can provide more tolerance when finding candidates. Similarly, finding candidates using the condition \textit{ANY} (instead of \textit{ALL}) also increases the tolerance. However, if the tagged entity contains stopwords such as “in”, “on”, etc., the pool of candidates will naturally grow very large and increase the computation cost for later part, because even completely irrelevant terms may contain stopwords. Therefore, we match based on non-stopwords only. To illustrate the points above, suppose a query is tagged with the entity 

\begin{Example}\label{ex1}
\centering
“severe burns on legs”,
\end{Example}

and one relevant term is “leg burn”. After stemming, “burns” and “legs” in Ex.\ref{ex1} become “burn” and “leg”, respectively, allowing "leg burn" to be considered as a candidate. Although the word “severe” is not in the term “leg burn”, the term is still considered a candidate because we selected using \textit{ANY}. The stopword “on” is ignored when finding candidate terms so that not every term that contains the word “on” is added to the candidate pool. When a candidate term, $C$, is found in this manner for the tagged entity, $E$, we calculate the semantic similarity score, $s$, between $C$ and $E$ in two steps. In the first step, calculate the maximum similarity score for each word in $C$ as shown in Figure \ref{fig:encoding}. Given a word in the candidate term, $C_i$ ($1 \leq i \leq m$, $m$ is the number of words in the candidate term) and a word in the tagged entity, $E_j$.Their similarity score, $s_{ij}$ (shown as the element in the boxed matrix in Figure \ref{fig:encoding}), is given by
\begin{equation}\label{eq3}
s_{ij} = 1-d(ELMo(C_i|C),ELMo(E_j|E))
\end{equation}
where $ELMo(C_i|C)$ and $ELMo(E_j|E)$ are the ELMo embeddings for the word $C_i$ and $E_j$, respectively. The ELMo embeddings have the same dimension for all words when using the same fine-tuned ELMo model. Thus, we can use a distance function (e.g., the cosine distance), denoted $d(\cdot)$ in equation \ref{eq3}, to compute the semantic similarity between words. 
In step 2, we calculate the candidate-entity relevance score (similarity) using the formula
\begin{equation}\label{eq4}
s(C,E) = \frac{1}{m} \sum_{i=1}^{m} max(\vec{S_i}) \cdot \mathbb{I} \{max(\vec{S_i}) > s_c\},  
\end{equation}
where $s_c$ is a score threshold, and $\mathbb{I} \{max(\vec{S_i}) > s_c\}$ is an indicator function that equals 1 if $max(\vec{S_i}) > s_c$ or equals 0 if not. In equation \ref{eq4} we define a metric that measures “information coverage” of the candidate terms with respect to a tagged entity. If the constituent words of a candidate term are relevant to the constituent words in the tagged entity, then the candidate term offers more information coverage. Intuitively, the more relevant words present in the candidate term, the more relevant the candidate is to the tagged entity. The purpose of the cutoff, $s_c$, is to screen the $(C_i,E_j)$ word pairs that are dissimilar, so that they do not contribute to information coverage. One can adjust the strictness of the entity - terminology matching by adjusting $s_c$. The higher we set $s_c$, the fewer candidate terms will be selected for a tagged entity. A normalization factor, $\frac{1}{m}$, is added to give preference to more concise candidate terms given the same amount of information coverage. 

We need to create an extra stopword list to include words such as “configuration” and “color”, and exclude these words from the word count for a candidate term. This is because the terms associated with the description of color or configuration usually have the word “color” or “configuration” in them. On the other hand, a user query normally does not contain such words. For example, a tagged entity in a user query could be “round yellow patches”, for which  the relevant terminologies include “round configuration” and “yellow color”. Since we applied a normalization factor, $\frac{1}{m}$, to the relevance score, the word “color” and “configuration” would lower the relevance score because they do not have a counterpart in the tagged entity. Therefore, we need to exclude them from word count. Once the process is complete, calculate $s(C,E)$ for all candidate terms and then we can apply a threshold on all $s(C,E)$ to ignore candidate terms with low information coverage. Finally, rank the terms by their $s(C,E)$ and return the ranked list as the results.

\section{Experiments}
\subsection{Data}
Despite the greater similarity between our task and the 2013 ShARe/CLEF Task 1, we use the clinical notes from the CE task in 2010 i2b2/VA on account of 1) the data from 2010 i2b2/VA being easier to access and parse, 2) 2013 ShARe/CLEF containing disjoint entities and hence requiring more complicated tagging schemes. The synthesized user queries are generated using the aforementioned dermatology glossary. Tagged sentences are extracted from the clinical notes. Sentences with no clinical entity present are ignored. 22,489 tagged sentences are extracted from the clinical notes. We will refer to these tagged sentences interchangeably as the i2b2 data. The sentences are shuffled and split into train/dev/test set with a ratio of 7:2:1. The synthesized user queries are composed by randomly selecting several clinical terms from the dermatology glossary and then combining them in no particular order. When combining the clinical terms, we attach the BIO tags to their constituent words. The synthesized user queries (13,697 in total) are then split into train/dev/test set with the same ratio. Next, each set in the i2b2 data and the corresponding set in the synthesized query data are combined to form a hybrid train/dev/test set, respectively. This way we ensure that in each hybrid train/dev/test set, the ratio between the i2b2 data and the synthesized query data is the same. 
 
\begin{figure*}[h]
\centering
\includegraphics[width=0.7\textwidth]{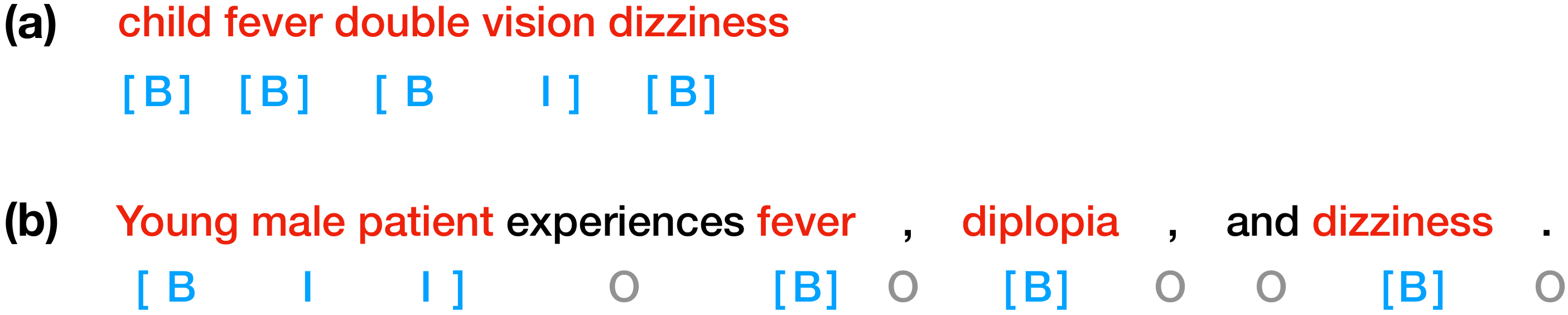}
\caption{(a) and (b) are equivalent in terms of what they describe. However, (a) is what a typical user query looks like and (b) is what a sentence in a clinical note looks like. Clinical entities, regardless of their type, are colored red. The tags for each token (word or punctuation) are labeled below the token. Notice that (a) is incoherent as a sentence and is several entities compounded together without any conjunction words or punctuation. (b) on the other hand, is cohesive. It contains a proportion of “O” tags. }
\label{fig:sample_query}
\end{figure*}
 
The reason for combining the two data is their drastic structural difference (See figure \ref{fig:sample_query} for an example). Previously, when trained on the i2b2 data only, the BiLSTM-CRF model was not able to segment clinical entities at the correct boundary. It would fail to recognize the user query in Figure \ref{fig:sample_query}(a) as four separate entities. On the other hand, if the model was trained solely on the synthesized user queries, we could imagine that it would fail miserably on any queries that resemble the sentence in Figure \ref{fig:sample_query}(b) because the model would have never seen an “O” tag in the training data.  Therefore, it is necessary to use the hybrid training data containing both the i2b2 data and the synthesized user queries.  

To make the hybrid training data, we need to unify the tags. Recall that in Section \ref{Intro} we point out that the tags are different for the different tasks and datasets. Since we use custom tags for dermatology glossary in our problem, we would need to convert the tags used in 2010 i2b2/VA. But this would be an infeasible job as we need experts to manually do that. An alternative is to avoid distinguishing the tag types and label all tags under the generic BIO tags. 

\subsection{Setup}
To show the effects of using the hybrid training data, we trained two models of the same architecture and hyperparameters. One model was trained on the hybrid data and will be referred to as hybrid NER model. The other model was trained on clinical notes only and will be referred to as i2b2 NER model. We evaluated the performance of the NER models by micro-F1 score on the test set of both the synthesized queries and the i2b2 data.

We used the BiLSTM-CRF implementation provided by the flair package \cite{akbik-flair}. We set the hidden size value to be 256 in the LSTM structure and left everything else at default values for the \textit{SequenceTagger} model on flair. For word embeddings, we used the ELMo embeddings fine-tuned on PubMed articles \footnote{\url{https://github.com/flairNLP/flair/blob/master/resources/docs/embeddings/ELMO_EMBEDDINGS.md}} and flair embeddings \cite{akbik-flair-embeddings} trained on $5\%$ of PubMed abstracts \footnote{\url{https://github.com/flairNLP/flair/blob/master/resources/docs/embeddings/FLAIR_EMBEDDINGS.md}}, respectively. We trained models for 10 epochs and experimented with different learning rate, mini batch size, and dropouts. We ran hyperparameter optimization tests to find the best combination. $S_c$ is set to be 0.6 in our experiment.

\subsection{Hyperparameter Tuning}
We defined the following hyperparameter search space:\\
\textit{embeddings}: [“ELMo on pubmed”, “stacked flair on pubmed”],\\
\textit{hidden\textunderscore size}: [128, 256],\\
\textit{learning\textunderscore rate}: [0.05, 0.1],\\
\textit{mini\textunderscore batch\textunderscore size}: [32, 64, 128].\\
The hyperparameter optimization was performed using Hyperopt \footnote{\url{https://github.com/hyperopt/hyperopt}}. Three evaluations were run for each combination of hyperparameters. Each ran for 10 epochs. Then the results were averaged to give the performance for that particular combination of hyperparameters.

\subsection{Results}
From the hyperparameter tuning we found that the best combination was\\ 
\textit{embeddings}: “ELMo on pubmed”, \\
\textit{hidden\textunderscore size}: 256, \\
\textit{learning\textunderscore rate}: 0.05, \\
\textit{mini\textunderscore batch\textunderscore size}: 32. \\

\begin{table}[h]
\centering
\begin{tabular}{ |c|c|c| } 
 \hline
 Model & Synthesized Queries & Clinical Notes \\ 
 \hline
 Hybrid NER & $0.995$ & $0.948$ \\ 
 \hline
 i2b2 NER & $0.441$ & $0.927$ \\ 
 \hline
\end{tabular}
\caption{Micro F1 scores of the two NER models two test sets.}
\label{table_1}
\end{table}

With the above hyperparameter setting, the hybrid NER model achieved a F1 score of $0.995$ on synthesized queries and $0.948$ on clinical notes while the i2b2 NER model achieved a F1 score of $0.441$ on synthesized queries and $0.927$ on clinical notes (See Table \ref{table_1}). 

\begin{figure}[h]
     \centering
     \begin{subfigure}[tb]{0.35\textwidth}
         \centering
         \includegraphics[width=\textwidth,height=6cm]{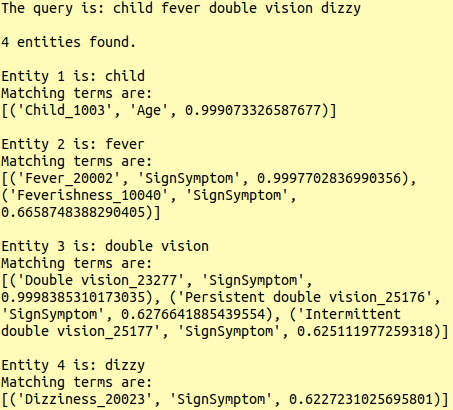}
         \caption{}
         \label{fig:result_a}
     \end{subfigure}
     \begin{subfigure}[tb]{0.35\textwidth}
         \centering
         \includegraphics[width=\textwidth,height=6cm]{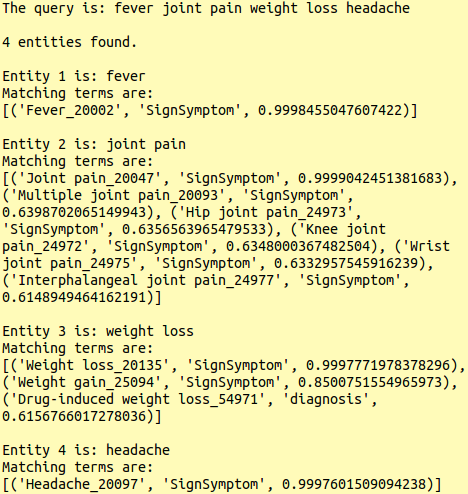}
         \caption{}
         \label{fig:result_b}
     \end{subfigure}
        \caption{The output of our framework. The query is shown in the first line. The tagged entities are shown one by one. The matched terms for an entity is shown in a list. For example, in (a) each matched term such as “('Child\textunderscore1003', 'Age', 0.999)” for entity 1 is shown as a tuple of three elements: the term and its cid ('Child\textunderscore1003'), the tag type of the term ('Age'), and the score of matching (0.999).}
        \label{fig:sample_results}
\end{figure}

Since there was no ground truth available for the retrieved terms, we randomly picked a few samples to assess its performance. Some example outputs of our complete framework on real user queries are shown in Figure \ref{fig:sample_results}. For example, from the figure we see that the query "child fever double vision dizzy" was correctly tagged with four entities: "child", "fever", "double vision", and "dizzy". A list of terms from our glossary was matched to each entity. In real world application, the lists of terms will be presented to the user as the retrieval results to their queries. 

\section{Discussion}
In most real user queries we sampled, the entities were tagged at the correct boundary and the tagging was complete (such as the ones shown in Figure \ref{fig:sample_results}). Only on a few user queries the tagging was controversial. For example, the query “Erythematous blanching round, oval patches on torso, extremities” was tagged as “Erythematous blanching” and “oval patches on torso”. The entity “extremities” was missing. The segmentation was also not correct. A more appropriate tagging would be “Erythematous blanching round, oval patches”, “torso”, and “extremities”. The tagging could be further improved by synthesizing more realistic user queries. Recall that the synthesized user queries were created by randomly combining terminologies from the dermatology glossary, which, while providing data that helped the model learn entity segmentation, did not reflect the co-occurrence information in real user queries. For example, there could be two clinical entities that often co-occur or never co-occur in a user query. But since the synthesized user queries we used combined terms randomly, the co-occurrence information was thus missing.

The final retrieval results of our framework were not evaluated quantitatively in terms of recall and precision, due the the lack of ground truth. When ground truth becomes available, we will be able to evaluate our framework more thoroughly. Recently, a fine-tuned BERT model in the medical domain called BioBERT \cite{Lee} has attracted some attention in the medical NLP domain. We could experiment with BioBERT embeddings in the future. We could also include query expansion technique for term matching. When finding candidate terms for an entity, our first step was still based on string matching. Given that there might be multiple entities that could be matched to the same term, it could be hard to include all these entities in the glossary and hard to match terms to these entities. 

\section{Conclusion}
In this project, we tackle the problem of extracting clinical concepts from user queries on medical search engines. By training a BiLSTM-CRF model on a hybrid data consisting of synthesized user queries and sentences from clinical note, we adopt a CE framework for clinical user queries with minimal effort spent on annotating user queries. We find that the hybrid data enables the NER model perform better on both tagging the user queries and the clinical note sentences. Furthermore, our framework is built on an easy-to-use deep learning NLP Python library, which lends it more prospective value to various online medical applications that employ medical search engines.    

\section*{Acknowledgment}
This paper results from a technical report of a project the authors have worked on with visualDx, a healthcare informatics company that provides web-based clinical decision support system. The authors would like to thank visualDx for providing them the opportunity to work on such an exciting project. In particular, the authors would like to thank Roy Robinson, the Vice President of Technology and Medical Informatics at visualDx, for providing the synthesized user queries, as well as preliminary feedback on the performance of our framework.

\bibliographystyle{unsrt}  
\bibliography{\BibDir/references}  


\end{document}